\documentclass[reprint, prl, aps, floatfix, twocolumn, showkeys]{revtex4-1}
\usepackage{times}
\usepackage{amssymb}
\usepackage{amsmath}
\usepackage{graphicx}
\usepackage{bm}
\usepackage{dcolumn}
\usepackage{color}
\usepackage{ulem}
\usepackage{float}
\usepackage{textcomp}
\restylefloat{table}
\usepackage{verbatim}
\usepackage{epstopdf}
\usepackage[colorlinks]{hyperref}
\usepackage[mediumspace,mediumqspace,squaren]{SIunits}

\begin{document}

\title{Anisotropic, two-dimensional, disordered Wigner solid}
\date{\today}

\author{Md.\ S. Hossain}
\author{M. K.\ Ma}
\author{K.\ A. Villegas-Rosales}
\author{Y. J.\ Chung}
\author{L. N.\ Pfeiffer} 
\author{K. W.\ West}
\author{K. W.\ Baldwin}
\author{M.\ Shayegan}
\affiliation{Department of Electrical Engineering, Princeton University, Princeton, New Jersey 08544, USA}

\begin{abstract}

The interplay between the Fermi sea anisotropy, electron-electron interaction, and localization phenomena can give rise to exotic many-body phases.  An exciting example is an \textit{anisotropic} two-dimensional (2D) Wigner solid (WS), where electrons form an ordered array with an anisotropic lattice structure. Such a state has eluded experiments up to now as its realization is extremely demanding: First, a WS entails very low densities where the Coulomb interaction dominates over the kinetic (Fermi) energy. Attaining such low densities while keeping the disorder low is very challenging.  Second, the low-density requirement has to be fulfilled in a material that hosts an anisotropic Fermi sea. Here, we report transport measurements in a clean (low-disorder) 2D electron system with anisotropic effective mass and Fermi sea. The data reveal that at extremely low electron densities, when the $r_s$ parameter, the ratio of the Coulomb to the Fermi energy, exceeds $\simeq 38$, the current-voltage characteristics become strongly nonlinear at small dc biases. Several key features of the nonlinear characteristics, including their anisotropic voltage thresholds, are consistent with the formation of a disordered, anisotropic WS pinned by the ubiquitous disorder potential.


\end{abstract}

\maketitle

Strong electron-electron interaction in clean two-dimensional electron systems (2DESs) leads to a plethora of many-body phases such as fractional quantum Hall liquid \cite{Tsui.PRL.1982}, Wigner solid (WS) \cite{Wigner.1934, Yoon.PRL.1999, Andrei.PRL.1988, Jiang.PRL.1990, Goldman.PRL.1990, Li.Sajoto.PRL.1991, Buhmann.PRL.1991, Santos.PRL.1992, Santos.PRB.1992, Yoon.PRL.1999, Chen.Nature.Phys.2006, Tiemann.Nat.Phys.2014, Jang.Nat.Phys.2016, Deng.PRL.2016, Deng.PRL.2019, Ma.PRL.2020, KAVR.PRR.2021, Hossain.Spin.Bloch.2020, Zhou.Nature.2021, Smole.Nature.2021, Falson.arxiv.2021}, and correlated magnetism \cite{Bloch.1929, Tanatar.1989, Attaccalite.PRL.2002, Drummond.PRL.2009, Sharpe.science.2019, Roch.PRL.2020, Polshyn.Nat.2020, Hossain.Spin.Bloch.2020, Hossain.preprint}.  Anisotropy introduces a new flavor to the interaction phenomena and triggers yet another set of unexpected correlated phases \cite{Wan.PRB.2002,  Zhou.2008, Gokmen.Natphy.2010, Xia.Natphys.2011, MNK.PRB.2011, Liu.PRB.2013, Yang.PRB.2012, Kamburov.PRL.2013, Kamburov.PRB.2014, Liu.PRL.2016,  Feldman.Science.2016, Jo.PRL.2017, Hossain.PRL.2018, Hossain.valley}, such as nematic quantum Hall states \cite{Xia.Natphys.2011, MNK.PRB.2011, Liu.PRB.2013, Yang.PRB.2012, Feldman.Science.2016,  Hossain.PRL.2018}.  A tantalizing example of Fermi sea anisotropy induced many-body states is a theoretically proposed anisotropic WS at zero magnetic field \cite{Wan.PRB.2002}.  This state is predicted to emerge at very low densities, harboring antiferromagnetic order \cite{Zhou.2008}.  However,  such a state at zero magnetic field has eluded experiments \cite{footnote.yang} because of the absence of a clean material platform where the low-density electrons occupy an anisotropic Fermi sea. 

Realization of a WS by itself is challenging. It requires Coulomb interaction ($E_C$) to dominate over the thermal energy ($k_BT$) and Fermi energy ($E_F$) \cite{Wigner.1934}. In a system where $E_F>>k_BT$ a \textit{quantum} WS should form when $E_C$ far exceeds $E_F$. This criterion can be realized at very low electron densities.  Quantitatively, in an ideal 2DES, Monte Carlo calculations \cite{Tanatar.1989, Attaccalite.PRL.2002, Drummond.PRL.2009} indicate that electrons should freeze into a quantum WS when the parameter $r_s$ exceeds $\simeq 35$; $r_s$ is the average inter-electron distance in units of the effective Bohr radius, equivalently, $r_s=E_C/E_F$. Achieving such large $r_s$ values and a sufficiently clean 2DES so that interaction phenomena are not completely hindered by excessive disorder and single-particle localization, however, is extremely challenging. Note that for a 2DES, $r_s= (m^* e^2/4 \pi \hbar^2 \varepsilon \varepsilon_0)/ (\pi n)^{1/2}$, where $m^*$ is the electron band effective mass, $\varepsilon$ is the dielectric constant, and $n$ is the 2DES density. In a GaAs 2DES ($m^*=0.067m_e$, where $m_e$ is the free electron mas and $\varepsilon=13$), $r_s\simeq35$ corresponds to a density of $n=2.5 \times 10^8$ cm$^{-2}$, which is indeed very difficult to attain \cite{Sajoto.PRB.1990, Zhu.PRL.2003}.  In systems with a larger $m^*$, high $r_s$ can be obtained at reasonable densities \cite{Yoon.PRL.1999, Hossain.Spin.Bloch.2020, Zhou.Nature.2021, Smole.Nature.2021, Falson.arxiv.2021}.  In fact, there are reports of a WS formation in GaAs 2D \textit{holes} \cite{Yoon.PRL.1999}, and AlAs \cite{Hossain.Spin.Bloch.2020}, MoSe$_{2}$ \cite{Zhou.Nature.2021, Smole.Nature.2021}, and ZnO \cite{Falson.arxiv.2021} electrons.  Here we attain extremely low densities and very large $r_s$ (up to $\simeq 50$) in an \textit{anisotropic} system and reveal an intriguing interplay between anisotropy and many-body localization in the WS regime.

\begin{figure*}[t!]
\includegraphics[width=0.98\textwidth]{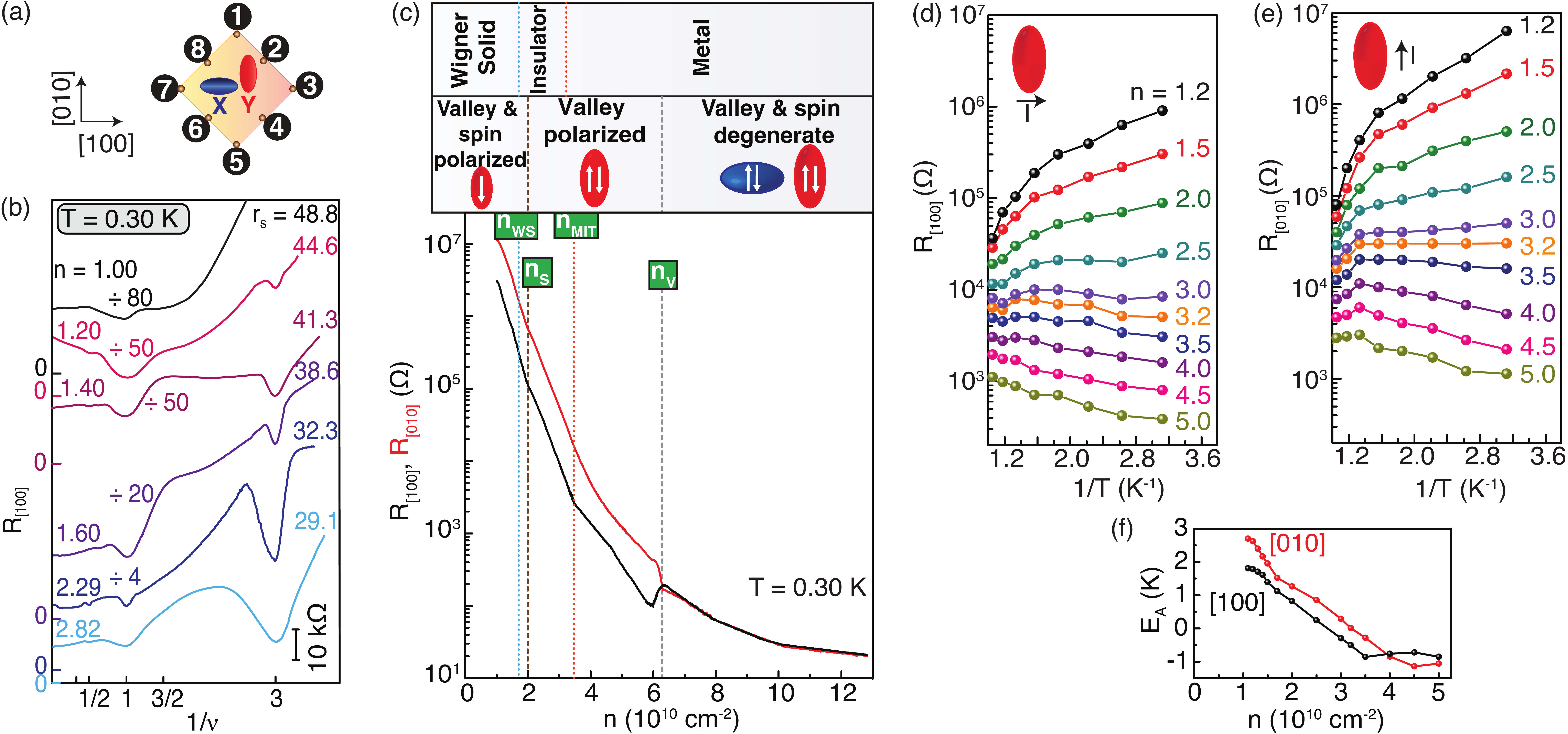}
\caption{\label{fig:fig1} Overview of our sample's geometry and its transport data. (a) Top view of our sample, an AlAs quantum well, in which the 2D electrons can occupy one or two anisotropic conduction-band valleys. Projections of the anisotropic Fermi seas of the X and Y valleys are shown in blue and red; these valleys have their major axes along the [100] and [010] crystallographic directions, respectively. The sample has a $\simeq 1.5 \times 1.5$ mm$^2$ van der Pauw geometry, and contacts to the sample are denoted by 1-8. For measuring $R_{[100]}$, we pass current from contact 8 to 2 and measure the voltage between contacts 6 and 4. For $R_{[010]}$, the current is passed from contact 8 to 6 and we measure the voltage between contacts 2 and 4. (b) Magnetoresistance traces taken along [100] as a function of the perpendicular magnetic field at different densities $n$, as indicated (in units of $10^{10}$ cm$^{-2}$), highlighting the extremely high quality of the 2DES at very low densities and large $r_s$. Traces are shifted vertically, and are plotted vs. $1/\nu$.  In all these traces, all the 2D electrons occupy the Y valley. (c) Resistances along [100] and [010] directions at zero strain as a function of density. $R_{[100]}$ and $R_{[010]}$ suddenly split when a spontaneous valley transition occurs at $n_V \simeq 6.3$. Also, two clear ``kinks” (marked with vertical lines) are seen at $n \simeq 3.5$ and 2.0 which closely correlate with the onset of the metal-insulator transition ($n_{MIT}$) and the spontaneous spin transition, respectively.  At even lower densities, below $n_{WS}\simeq 1.80$, we start to see non-linear $I$-$V$ characteristics, suggestive of a pinned WS; this is the main subject of our work presented here. (d,e) Arrhenius plots of $R_{[100]}$ and $R_{[010]}$. Both $R_{[100]}$ and $R_{[010]}$ exhibit metallic transport for $n \gtrsim 3$, and an insulating behavior for smaller $n$. $R_{[100]}$ and $R_{[010]}$ exhibit approximately an activated behavior at low $T$ ($\lesssim 0.6$ K), $R \propto e^{E_A/2k_BT}$, for $n \lesssim 3 $. (f) shows the deduced $E_A$ along [100] and [010] vs. density. }
\end{figure*}

\begin{figure}[t!]
\includegraphics[width=0.49\textwidth]{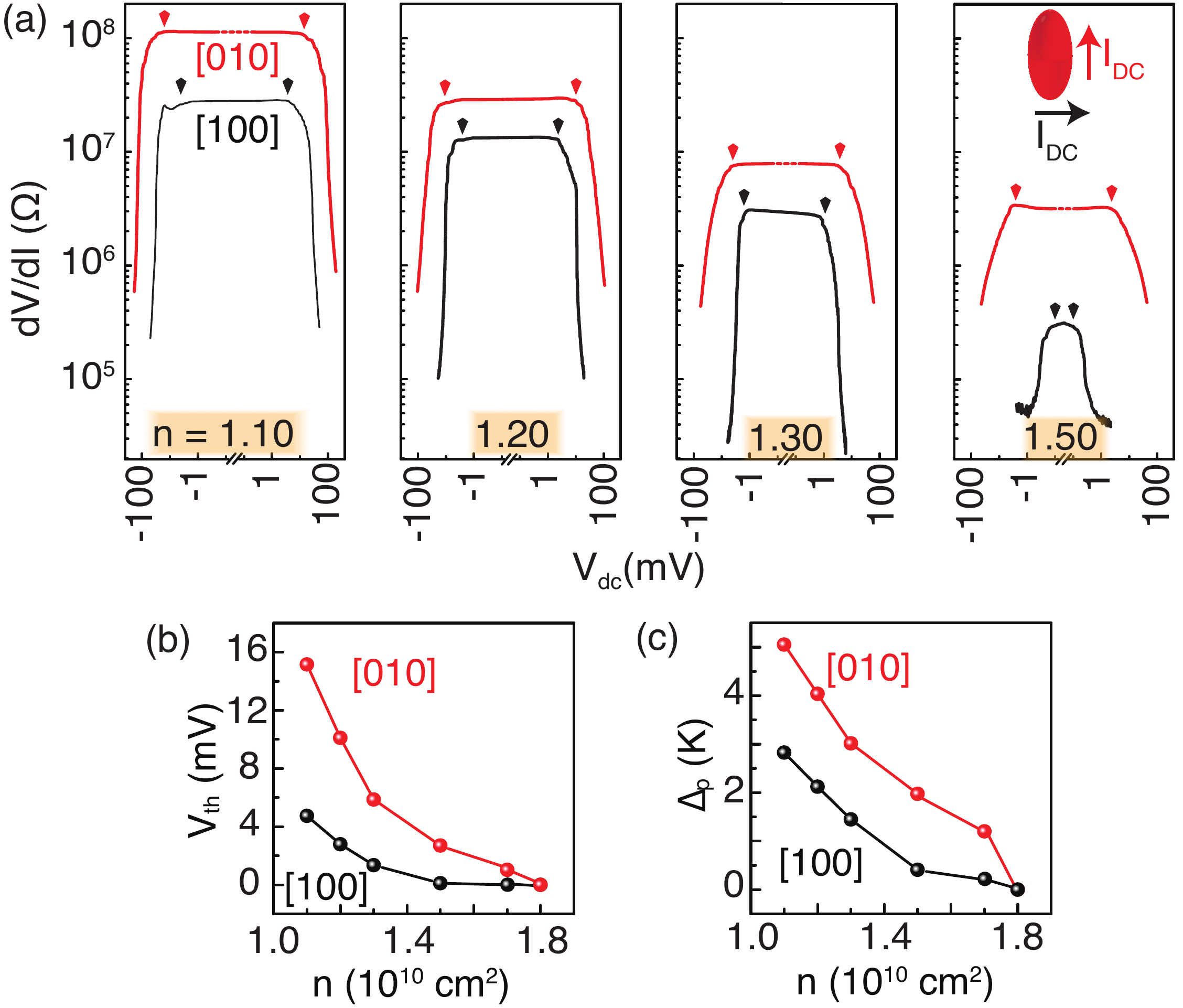}
\caption{\label{fig:fig2a} (a) $dV/dI$ data at $T=0.30$ K along the [100] (black traces) and [010] (red traces) crystallographic directions, plotted as a function of the dc voltage drop along the sample; data are shown for $n=1.10$, 1.20, 1.30,  and 1.50 (in units of $10^{10}$ cm$^{-2}$). Along both directions $dV/dI$ traces show a reasonably abrupt drop above a threshold voltage ($V_{th}$). The vertical arrows (placed symmetrically with respect to $V_{dc} = 0$) mark $V_{th}$, and are based on the average value of $V_{th}$ for $+V_{dc}$ and $-V_{dc}$. (b) Extracted $V_{th}$ and (c) pinning gap ($\Delta_p$) plotted vs. $n$, highlighting the increase in $V_{th}$ and $\Delta_p$ as $n$ is lowered. }
\end{figure}

\begin{figure}[t!]
\includegraphics[width=0.49\textwidth]{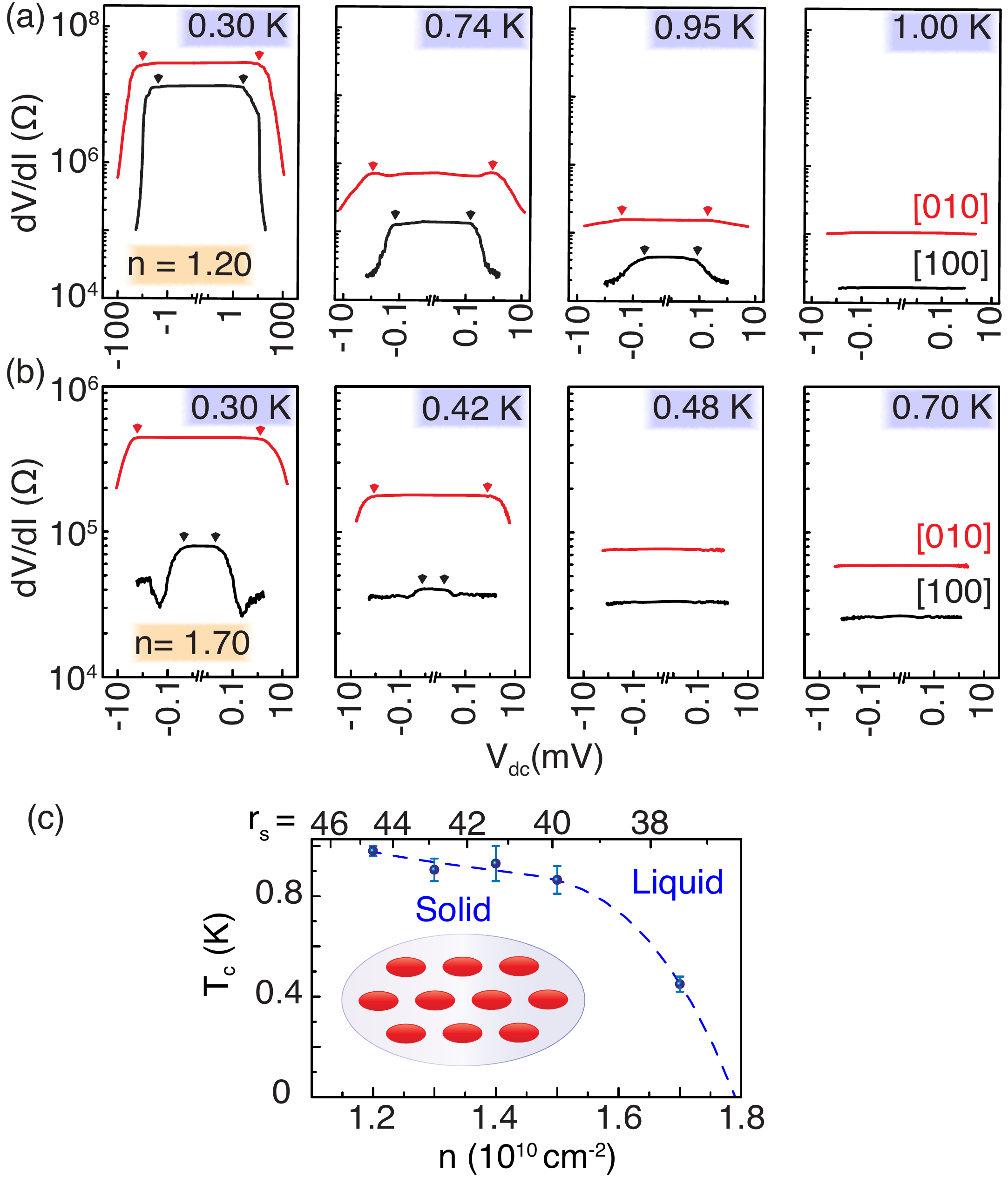}
\caption{\label{fig:fig3} Temperature dependence of $dV/dI$ vs. $V_{dc}$ at (a) $n=1.20$ and (b) $n=1.70$, measured along [100] and [010] directions.  (c) Critical temperature ($T_c$), below which the $I$-$V$ characteristic becomes nonlinear, plotted vs. electron density. The data suggest a drop in $T_c$ with increasing density, extrapolating to 0 K as the onset density for the pinned WS formation ($n \simeq 1.80$) is reached. The dashed curve is a guide to the eye. The inset displays schematically an anisotropic WS; the shaded regions represent the electron charge distribution with anisotropic Bohr radii, reflecting the anisotropic electron effective mass.  }
\end{figure}

Our material platform is a high-quality 2DES confined to a 21-nm-wide AlAs quantum well \cite{Shayegan.AlAs.Review.2006, Lay.APL.1993, Depoortere.APL.2002, Chung.PRM.2018, SM}. The 2D electrons in the AlAs well occupy two in-plane conduction-band valleys with longitudinal and transverse effective masses $m_l = 1.1$, and $m_t=0.20$, leading to an effective in-plane $m^*=(m_l m_t)^{1/2}=0.46$ \cite{Shayegan.AlAs.Review.2006, Lay.APL.1993, Depoortere.APL.2002, Chung.PRM.2018, SM}.  Via applying uniaxial strain along the [100] direction we transfer all the electrons to the valley whose longer Fermi wavevector axis is along the $[010]$ direction \cite{Shayegan.AlAs.Review.2006}; we refer to this as the Y valley, see Fig. 1(a). As discussed below, this single-valley occupancy can also be achieved, in the absence of uniaxial strain, by tuning $n$ to very small values; $n$ is controlled \textit{in situ} by applying voltage to a back gate.  The exceptionally high quality of our 2DES can be inferred from Fig. 1(b) which shows magnetoresistance traces measured along [100] as a function of the perpendicular magnetic field ($B$) for several $n$ ranging from $2.82$ to $1.00$, in units of $10^{10}$ cm$^{-2}$ which we use throughout this paper. We plot the data as a function of $\nu^{-1}$ so that the quantum Hall states line up for the different traces; $\nu=nh/eB$ is the Landau level filling factor. The traces exhibit clear minima at $\nu=1/3$ down to $n=1.20$ ($r_s=44.6$ \cite{footnote.mass}), indicating the presence of fractional quantum Hall states and attesting to the very high sample quality.

Figure 1(c), which presents the sample resistance along [100] and [010], provides an overview of the different transitions and phases observed in our sample as $n$ is lowered. Here there is no uniaxial strain applied so that the two in-plane valleys (X and Y), with their major axes along [100] and [010], are equally occupied at high densities; $R_{[100]}$ and $R_{[010]}$ are essentially equal and the 2DES is isotropic. As $n$ is lowered below $\simeq 6.3$ ($r_s \simeq 20$), all the electrons suddenly transfer to the Y valley, leading to $R_{[010]} > R_{[100]}$ (note that the electrons in the Y valley have a larger effective mass along [010] compared to [100]). This abrupt, spontaneous valley-polarization, which is akin to Bloch spin ferromagnetism transition, has been discussed in detail elsewhere \cite{Hossain.valley}; see also \cite{Ahn.Das.Sarma.Valley.cond.mat.2021}.  At lower densities, when $n \lesssim 3.2$ ($r_s  \gtrsim 28$), the 2DES turns insulating as signaled by the temperature dependence of $R_{[100]}$ and $R_{[010]}$ at low temperatures; see Figs. 1(d,e). Note in Fig. 2(d) that near this transition to an insulating phase, we observe a ``kink" in resistance vs. $n$ traces. At even lower densities, $n \lesssim 2.0$ ($r_s  \gtrsim 35$), the 2DES makes yet another transition to a fully-spin-polarized state \cite{Hossain.Spin.Bloch.2020}. Finally, at $n \lesssim 1.8$ ($r_s  \gtrsim 37$), the 2DES becomes highly insulating while still maintaining its anisotropic transport coefficients, and develops a strongly nonlinear \textit{I-V} characteristic at low temperatures \cite{footnote.kink}. This phase is the focus of our report here. 


The highlight of our study is presented in Fig. 2(a) where we show the differential resistance $dV/dI$ along [100] and [010] vs. the dc bias voltage ($V_{dc}$) at very low densities. For Fig. 2 data, as well as in Fig. 3, the 2D electrons occupy only the Y valley. We use log-log plots to cover several orders of magnitude in $dV/dI$ and $V_{dc}$. The data in Fig. 2(a) show nonlinear $I$-$V$ behavior, with clear threshold voltages ($V_{th}$) that are larger along [010] compared to [100] (for a given density) \cite{heating}. We attribute these nonlinearities to the formation of a WS pinned by the ubiquitous disorder potential. Similar nonlinearities at small biases have been observed in previous studies of dilute GaAs 2D holes at $B=0$ \cite{Yoon.PRL.1999}, and at very small Landau level fillings in the extreme quantum limit \cite{Goldman.PRL.1990, Li.Sajoto.PRL.1991}, which also concluded that the nonlinearity in $I$-$V$ was a signature of the depinning of a WS. The conclusion for high-$B$ data has been further corroborated by numerous experimental studies at very small Landau level fillings, including transport \cite{Jiang.PRL.1990}, noise \cite{Li.Sajoto.PRL.1991}, magneto-optics \cite{Buhmann.PRL.1991}, microwave \cite{Andrei.PRL.1988, Chen.Nature.Phys.2006}, nuclear magnetic resonance \cite{Tiemann.Nat.Phys.2014}, tunneling \cite{Jang.Nat.Phys.2016}, geometric resonance \cite{Deng.PRL.2016}, and capacitance \cite{Deng.PRL.2019} measurements. Therefore, it is likely that in our 2DES, too, the nonlinear \textit{I-V} signals the formation of a pinned WS at $r_s  \gtrsim 37$ ($n \lesssim 1.80$). Below we elaborate on several important features of the nonlinearities in Fig. 2(a), and argue that they are consistent with an interpretation of our data in terms of a disordered, \textit{anisotropic} WS.

A key observation in our data is that the magnitude of the threshold electric field ($E_{th}$) is consistent with the expected depinning of a 2D WS \cite{Yoon.PRL.1999}. For example, Chui \cite{Chui.PLA.1993} proposed a model in which the WS is weakly pinned by the potential of the remote ionized dopants, and transport takes place through the creation of dislocation pairs and quantum tunneling. In this model, it is estimated that:
\begin{equation}\label{Eq1}
E_{th} \simeq 0.09n_{i} a^2 e / (4 \pi \varepsilon d^{3}),
\end{equation}
where $n_i$ is the density of remote ionized impurities, $a=(\pi n)^{-1/2}$, and $d$ is the spacer-layer thickness.  Using relevant and reasonable parameters for our sample ($\varepsilon=10$, $d=68$ nm, $n=1.3$, and $n_i = 2 \times 10^{11}$ cm$^{-2}$ \cite{Footnote.A}), we estimate $E_{th} \simeq 5$ V/m. Given that the distance between voltage contacts in our sample [e.g., contacts 2 and 8 in Fig. 1(a)] is $\simeq 1$ mm, $E_{th} \simeq 5$ V/m corresponds to $V_{th} \simeq 5$ mV; this is of the order of $V_{th}$ we observe in our experiments [see Fig. 2(b)]. Given the uncertainty in the value of $n_i$, we believe our data are consistent with the model of Ref. \cite {Chui.PLA.1993} for a pinned WS. 

A related implication of Fig. 2 data is the WC pinning energy $\Delta_p \simeq \hbar (eE_{th}/m^*a)^{1/2}$ that can be deduced from the measured $V_{th}$ \cite{Yoon.PRL.1999}. In Fig. 2(c) we plot, as a function of $n$, the values of $\Delta_p$ deduced from the V$_{th}$ data of Fig. 2(b). Note that Eq. (1) and the above expression for $\Delta_p$ imply that both V$_{th}$ and $\Delta_p$ should increase as the density is lowered. This behavior is seen qualitatively in Figs. 2(b,c). Moreover, the magnitude of $\Delta_p$ should be comparable to the activation energies extracted from the Arrhenius plots in Fig. 1(f), and this is indeed true, as both sets of data yield energies of a few K.  

The V$_{th}$ data can be used to also extract an approximate size for the WS domain: $L_0^2 = 0.02 e / 4 \pi \varepsilon \varepsilon_0 E_{th}$ \cite{Li.Sajoto.PRL.1991}.  This expression is based on a theory by Fukuyama and Lee \cite{Fukuyama.PRB.1978}, and uses the 2D WS shear modulus calculated by Bonsall and Maradudin  \cite{Bonsall.PRB.1977}. Evaluating this expression for our data at $n=1.3$, where $E_{th} \sim4$ V/m [see Fig. 2(b)], we find $L_0^2 \sim 7 \times 10^{-13}$ m$^2$, implying that a typical WS domain contains $n L_0^2 \sim  90$ electrons. This is admittedly not a very large domain, but is plausible, given the amount of disorder in our sample. Note also that, in this model, the anisotropy of $E_{th}$ implies that the WS domains are anisotropic too, with the domain size along [010] being larger than along [100].

Data of Fig. 3 illustrate another aspect of our 2DES. In Figs. 3(a,b) we show the temperature dependence of nonlinear $dV/dI$ along [100] and [010] at two different densities, $n=1.20$ and 1.70. The data indicate that, for a given $n$ and direction, $V_{th}$ moves to smaller values as $T$ is raised, and the nonlinearity disappears above a critical temperature $T_c$.  We interpret this disappearance to signal the thermal melting of the WS. As seen in Fig. 3(a), $T_c$ at $n=1.20$ appears to be between 0.95 K and 1.00 K.  Note that the data along [100] and [010] exhibit a similar temperature dependence, implying the same $T_c$ along both directions. When we raise the density, $T_c$ decreases. For example, at $n=1.70$, the nonlinearity disappears for $T \gtrsim 0.48$ K [Fig. 3(b)]. Figure 3(c) summarizes the measured $T_c$ as a function of $n$. As we raise the density, up to $n=1.50$, $T_c$ decreases only slightly. However, $T_c$ drops rapidly at higher $n$, implying a quantum melting of the WS at $n \simeq 1.80$.

The critical temperatures we summarize in Fig. 3(c) are about an order of magnitude larger than the melting temperatures theoretically estimated for an ideal (zero-disorder) quantum WS with parameters similar to our 2DES [see, e.g. Fig. 2(a) of Ref. \cite{Das.Sarma.PRB.2003} for GaAs 2D holes]. Also, in our sample the Fermi energy at the lowest densities is small and becomes comparable to the measurement temperatures; this implies that our 2DES approaches the classical limit. It is likely that pinning by the impurities at very low densities is what leads to the much larger melting temperatures in our sample. Indeed, recent theoretical studies, partly motivated by our results, indicate that the effective melting temperature of a WS in the presence of disorder can be significantly enhanced relative to the pristine case as the disordered WS becomes fragmented and localized by the disorder potential \cite{VuDasSarma.cond.mat.2021, Shklovskii.Unpublished.2021}.  Moreover, in a system with disorder, the 2DES eventually attains an Anderson localized state in the limit of extremely small densities because the disorder potential would dominate over the ever-decreasing Coulomb repulsion energy. We note that $T_c$ in Fig. 3(c) are comparable to the melting temperatures reported for the magnetic-field-induced WS phases of dilute 2DESs in GaAs \cite{Goldman.PRL.1990, Chen.Nature.Phys.2006, Deng.PRL.2019} and AlAs \cite{KAVR.PRR.2021}, and 2D holes in GaAs \cite{Ma.PRL.2020}. They are much smaller than those recently quoted for the zero-field WS phases in MoSe$_2$ \cite{Zhou.Nature.2021, Smole.Nature.2021}, but much larger than in ZnO \cite{Falson.arxiv.2021}. 

We close by making several comments concerning the anisotropy observed in our data. First, Fig. 1(b) data show that the transport anisotropy persists in the entire low-density range once all the electrons are transferred to the Y valley (below $n=6.3$). This includes the high densities (e.g., $n \simeq 4$), when the 2DES is in a ``metallic" state, and also deep into the insulating phases at very low densities. Second, $V_{th}$ for the onset of drops in $dV/dI$ vs. $V_{dc}$ also show an anisotropy, with $V_{th}$ along [010] being larger compared to [100] (Figs. 2 and 3). These features are very likely related to the anisotropic effective mass in our 2DES. For a 2DES like ours, with a large effective mass along the $y$ direction ($m_l=1.1$) and a small mass along $x$ ($m_t=0.20$), the electron charge distribution has an elliptical shape which is rotated by 90$^o$ with respect to the shape of the Fermi sea/contour [see Fig. 3(c) inset]. Such a charge distribution would imply that transport along $y$ would be harder compared to $x$ as the electrons are more localized (smaller Bohr radius) and have less overlap along $y$. This is consistent with our observations. Now, theory \cite{Wan.PRB.2002} suggests that in a 2DES with an anisotropic effective mass, the WS is distorted from its conventional, triangular lattice, and has a shape schematically shown in Fig. 3(c) inset. Note that, compared to the conventional triangular lattice, here the lattice is stretched along the $x$ axis. (The stretch along the axis of the smaller mass reduces the longitudinal energy of lattice vibrations, thus lowering the ground state energy \cite{Wan.PRB.2002}). It then appears that the transport and $V_{th}$ anisotropies would depend on the competition between the anisotropies of the electron charge distribution and the WS lattice, e.g., a very large elongation of the lattice might eventually lead to a reversal of the transport anisotropy, with resistance and $V_{th}$ along $x$ exceeding those along $y$. We do not observe such reversal of anisotropies, suggesting that the transport anisotropy in our 2DES is controlled by the electron charge distribution anisotropy.

\begin{acknowledgments}
We acknowledge support through the U.S. Department of Energy Basic Energy Sciences (Grant No. DEFG02-00-ER45841) for measurements, and the National Science Foundation (Grants No. DMR 2104771, No. ECCS 1906253, DMR 1709076, and MRSEC DMR 2011750), the Eric and Wendy Schmidt Transformative Technology Fund, and the Gordon and Betty Moore Foundation’s EPiQS Initiative (Grant No. GBMF9615 to L. N. P.) for sample fabrication and characterization. We also thank J. K. Jain, S. Das Sarma, L. W. Engel, D. A. Huse, M. A. Mueed, and B. Shklovskii for illuminating discussions.
\end{acknowledgments}

\bibliographystyle{apsrev}

\begin{thebibliography}{99}
\bibitem{footnote.mass} To calculate $r_s$ values, we use the effective in-plane band mass $m^*$ which is equal to $(m_l m_t)^{1/2}=0.46$. When electrons occupy a single valley,  one can define $r_s$ along [100] and [010] separately. This would lead to an anisotropic $r_s$. For example, when electrons occupy only the Y valley,  $m_{[100]} =m_t = 0.2$ and $m_{[010]} =m_l = 1.1$. Therefore, $r_{s[010]} =5.5 \times r_{s[100]}$.  $r_{s[100]}$ and $r_{s[010]}$ are related to the $r_s$ values that we used as follows: $r_s/0.46 = r_{s[100]}/0.2 = r_{s[010]}/1.1$.
\bibitem{footnote.kink}Note in Fig. 1(c) that near $n \simeq 2$, $R_{[100]}$ and $R_{[010]}$ show a second ``kink" towards higher resistances at lower temperatures.
\bibitem{heating} Note that the power dissipated in the sample is much smaller than the cooling power of the cryostat. Also, Joule self-heating in the 2DES is not responsible for the nonlinearities we report: At a given density and temperature, the power required to achieve the same change in the differential resistance is more than an order of magnitude different along the two directions. We therefore conclude that the effect of heating is negligible in our measurements.
\bibitem{Footnote.A}  The parameter $n_i$ is not precisely known for modulation-doped samples; it is typically smaller than the concentration of the $\delta$-doping layer but larger than the 2DES density in the quantum well.
\end{thebibliography}


\end{document}